\documentclass[sigconf,screen]{acmart}

\usepackage{color}
\usepackage{tabularray}
\usepackage{multirow}
\usepackage{natbib}
\usepackage{graphicx}
\usepackage{subfigure}
\usepackage{textcomp}
\usepackage{xcolor}
\usepackage{algpseudocode}
\usepackage{algorithm}
\theoremstyle{plain}

\theoremstyle{definition}

\theoremstyle{remark}

\usepackage{graphicx}
\usepackage{tabularx}
\usepackage{amsmath,xparse,mleftright}
\usepackage[export]{adjustbox}
\usepackage{balance}
\usepackage{tcolorbox}

\usepackage{enumitem}
\usepackage{xcolor}
\usepackage{pifont}
\usepackage{xspace}

\usepackage{setspace}

\newcommand{\ie}{\emph{i.e.},\xspace}
\newcommand{\etal}{\emph{et al.},\xspace}

\newcommand{\name}{\textsc{TraceDiag}\xspace}

\newcommand{\traceark}{\textsc{TraceArk}\xspace}
\newcommand{\causalrca}{\textsc{CausalRCA}\xspace}

\newcommand\boxwidth{8.5cm}
\newcommand\innerwidth{2mm}

\newcounter{insightC}

\newtcolorbox{mybox}[2][]{colbacktitle=red!10!white, colback=gray!10!white,coltitle=gray!70!black, title={#2},fonttitle=\bfseries,#1}

\setcopyright{acmlicensed}
\acmPrice{15.00}
\acmDOI{10.1145/3611643.3613864}
\acmYear{2023}
\copyrightyear{2023}
\acmSubmissionID{fse23industry-p12-p}
\acmISBN{979-8-4007-0327-0/23/12}
\acmConference[ESEC/FSE '23]{Proceedings of the 31st ACM Joint European Software Engineering Conference and Symposium on the Foundations of Software Engineering}{December 3--9, 2023}{San Francisco, CA, USA}
\acmBooktitle{Proceedings of the 31st ACM Joint European Software Engineering Conference and Symposium on the Foundations of Software Engineering (ESEC/FSE '23), December 3--9, 2023, San Francisco, CA, USA}
\received{2023-05-18}
\received[accepted]{2023-07-31}

\begin{document}

\title{\name: Adaptive, Interpretable, and Efficient Root Cause Analysis on Large-Scale Microservice Systems
}

\settopmatter{authorsperrow=4} 

\author{Ruomeng Ding}
\affiliation{%
\institution{
     Microsoft
 }
 \country{China}
}

\author{Chaoyun Zhang, \\Lu Wang, Yong Xu}
\affiliation{%
\institution{
     Microsoft
 }
 \country{China}
}

\author{Minghua Ma, \\Xiaomin Wu}
\affiliation{%
\institution{
     Microsoft
 }
 \country{China}
}

\author{Meng Zhang, \\Qingjun Chen}
\affiliation{%
\institution{
     Microsoft 365
 }
 \country{China}
}

\author{Xin Gao,\\Xuedong Gao, Hao Fan}
\affiliation{%
\institution{
     Microsoft 365
 }
 \country{China}
}

\author{Saravan Rajmohan}
\affiliation{%
\institution{
     Microsoft 365
 }
 \country{USA}
}

\author{Qingwei Lin}
\affiliation{%
\institution{
     Microsoft
 }
 \country{China}
}

\author{Dongmei Zhang}
\affiliation{%
\institution{
     Microsoft
 }
 \country{China}
}




\renewcommand{\shortauthors}{Ding \etal}


\begin{abstract}
Root Cause Analysis (RCA) is becoming increasingly crucial for ensuring the reliability of microservice systems. However, performing RCA on modern microservice systems can be challenging due to their large scale, as they usually comprise hundreds of components, leading significant human effort. This paper proposes \name, an end-to-end RCA framework that addresses the challenges for large-scale microservice systems. It leverages reinforcement learning to learn a pruning policy for the service dependency graph to automatically eliminates redundant components, thereby significantly improving the RCA efficiency. The learned  pruning policy is interpretable and fully adaptive to new RCA instances. With the pruned graph, a causal-based method can be executed with high accuracy and efficiency. The proposed \name framework is evaluated on real data traces collected from the Microsoft Exchange system, and demonstrates superior performance compared to state-of-the-art RCA approaches. Notably, \name has been integrated as a critical component in the Microsoft M365 Exchange, resulting in a significant improvement in the system's reliability and a considerable reduction in the human effort required for RCA.
\end{abstract}




\maketitle
\vspace*{-0.5em}
\section{Introduction}
The microservices architecture has gained widespread popularity in modern production systems that leverage cloud technologies \cite{soldani2022anomaly}. This approach involves decomposing large systems into smaller, self-contained components that communicate with each other, thereby enabling agile development and deployment \cite{liu2019jcallgraph, gan2019seer}. For instance, Twitter has about 1,200 microservices on the server side \cite{musk} and Microsoft Exchange employs hundreds of microservices to enable email delivery, with each microservice interacting with others based on complex dynamic dependencies \cite{TraceArk}. In such intricate systems, anomalous incidents may occur in individual components, thereby impacting downstream components and resulting in unexpected end-to-end latency and a high number of delayed email deliveries \cite{liu2021microhecl}. These issues can significantly harm the customer experience, leading to substantial financial losses \cite{ma2022empirical}, and adversely affect the reputation of service providers.

To ensure the reliability of cloud services, on-call engineers  rely on a range of system data sources such as key performance indicators \cite{ma2018robust}, logs \cite{liu2022uniparser}, and traces \cite{TraceArk} to develop automated and intelligent incident detection and root cause analysis (RCA) solutions \cite{zhou2018fault, TraceArk, zhang2021halo, ma2020diagnosing, wang2023rca}. The goal is to quickly mitigate incidents and minimize their impact on business operations by analyzing anomaly propagation chains to pinpoint the root causes. However, this is a challenging task, particularly given the complex interactions and interdependencies among the system components, which is further compounded by the scale of modern cloud systems. Research shows that RCA can take several hours on average without the use of automated tools, making it a time-consuming and arduous process \cite{liu2021microhecl, wang2018cloudranger}. In large systems like Microsoft Exchange, where service performance would be influenced by multiple factors including infrastructure scale, automatic scaling, dynamic load balancing, system updates, and security measures \cite{budhathoki2021did, guo2020graph}, RCA becomes even more intricate.

\begin{figure}[t]
\centering
\includegraphics[width=0.9\columnwidth]{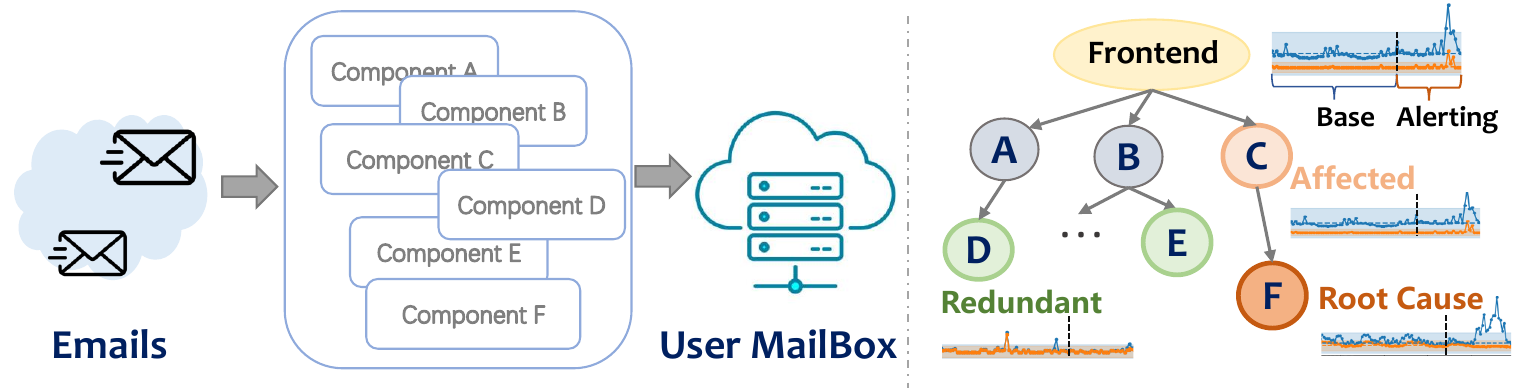}
\vspace*{-1.5em}
\caption{An example of trace in Microsoft Exchange system.
\label{fig:intro}}
\vspace*{-2.5em}
\end{figure}

Traces serve as a crucial component of system observability, capturing microservice invocation dependency and associated performance metrics. Graph analysis techniques are commonly employed in trace analysis to automatically localize faults by traversing service dependency graphs and detecting anomalies based on service quality metrics~\cite{zhou2019latent, wang2019grano, brandon2020graph, guo2020graph}. However, the scale of modern microservice systems can lead to challenges in accurate anomaly detection and efficient graph traversal, as they generate massive volumes of system traces containing hundreds to thousands of components and their dependencies. One contributing factor to the complexity of RCA is the presence of redundant components that have no impact on the incident and can be safely removed without compromising the RCA performance~\cite{yu2023cmdiagnostor, liu2021microhecl}. Fig.~\ref{fig:intro} illustrates an example trace, depicting the latency evolution of each component. The root cause component $F$ experiences a surge in latency, subsequently affecting component $C$ and the frontend node. However, other components that do not interact with $F$ maintain normal latency behavior and do not contribute to the RCA process. These components can be considered ``redundant'' and eliminating them from the trace can reduce noise and enhance the efficiency of RCA.

Pruning redundant components from dependency graphs is a widely adopted technique in the RCA process \cite{liu2021microhecl, yu2023cmdiagnostor}. However, determining which components to prune poses a significant challenge. Aggressive pruning, where too many components are removed, runs the risk of accidentally removing the root cause node, rendering further RCA impossible. Conversely, a lenient pruning strategy that preserves too many components can result in inefficient RCA and generate noisy outcomes. Current pruning policies are often impromptu, based on engineers' expertise and tailored to specific incidents, lacking interpretability and susceptible to system changes over time. Consequently, significant human effort is required, and a lack of standardized and unified rules persists. Therefore, there is a pressing need for an adaptive service graph pruning policy that can offer interpretability in the pruning process. 

To address the aforementioned challenges, this paper introduces \name, a novel framework for RCA in large-scale microservice systems. \name leverages reinforcement learning (RL) to acquire an automated, interpretable, and adaptable pruning policy that effectively removes redundant components, enhancing the efficiency and accuracy of the RCA process. The policy is based on graph pruning rules derived from experienced engineers and comprehensive trace analysis, ensuring domain knowledge and industry best practices are incorporated. The pruned service graph is then utilized for RCA using causal methods. The practicality and effectiveness of \name have been demonstrated through its deployment as a core RCA engine in a Microsoft microservice system. In summary, the contributions of this paper include:

\begin{itemize}[leftmargin=*]
    \item We introduce \name, an end-to-end RCA framework for large-scale microservice systems. The \name integrates graph pruning to enhance RCA efficiency and a causal RCA method that surpasses correlation-based approaches to provide more accurate and robust results.
    \item We employ RL to automate the graph pruning by selecting from a pre-defined pruning action pool designed through comprehensive trace analysis. This ensures interpretability and adaptability to new RCA instances.
    \item We evaluate our proposed framework on real data traces collected from the Microsoft Exchange system, demonstrating superior performance compared to state-of-the-art RCA approaches.
    \item The proposed framework has been integrated as a critical component in the Microsoft M365 Exchange system, leading to 75.1\% higher RCA accuracy and reduce time for RCA by up to 96.5\%. This yields a significant improvement in the system's reliability and considerable reduction for the human effort for RCA.
\end{itemize}
To the best of our knowledge, our \name is the first to apply RL as the trace pruning process for the RCA of microservice systems.

\vspace*{-0.5em}
\section{Background and Motivation}
This section provides an overview of the Microsoft Exchange Microservice System, which serves as the experiment and implementation of the \name approach. We also formulate the problem of trace-based RCA in this section.

\vspace*{-0.5em}
\subsection{Microsoft Exchange Microservice System}
Microsoft Exchange is a cloud-based email delivery system that offers an efficient and reliable message routing and delivery service. Email delivery latency is a crucial performance metric that constitutes the primary service level agreement provided to customers. The message delivery process in the Microsoft Exchange cloud system involves numerous components from various servers and services, processing billions of emails daily. We illustrate the system in the left of Fig.~\ref{fig:intro}.

To analyze email delivery latency in Microsoft Exchange, an email request generates a trace that captures its invocation dependencies. The trace is represented as a directed tree structure, denoted as $\mathcal{G}$, where nodes correspond to microservices or functions invoked, and edges represent their interdependent relationships \cite{TraceArk}. Due to the interdependent nature of the edges in $\mathcal{G}$, they can depict the causal relationship between microservices, which goes beyond simple correlation. The trace also records relevant information, such as structural attributes, service metadata, call latency, and temporal information, for subsequent analysis. The trace data, coupled with the latency evolution of each microservice, constitute multivariate time series \cite{zhang2020microscope, zhang2017zipnet, zhang2021cloudlstm, zhang2019deep} with a structural topology \cite{yu2018spatio, ali2022exploiting, mohamed2020social}. Note that in contrast to previous studies such as \cite{pan2021faster, meng2020localizing}, we have access to and knowledge of the trace topology. As a result, there is no need for the causal discovery \cite{granger1969investigating, long2021air, wu2020connecting}  step to construct the service dependency graph in this work.

Each email request incurs end-to-end latency that includes delays from all the microservices in the trace. If the latency exceeds the established service level agreement (SLA), it can result in slow email deliveries, leading to customer dissatisfaction and escalation. To detect anomalous latency surges, Microsoft Exchange employs \traceark \cite{TraceArk}, and subsequently performs RCA to localize problematic components that contribute to the anomaly. RCA plays a critical role in identifying the cause of the issue, enabling prompt mitigation actions, and minimizing the impact on business.

\vspace*{-1em}
\subsection{Root Cause Analysis on Trace}
Incidents occurring in the microservice system frequently lead to a considerable number of emails that fail to meet the latency Service Level Agreement (SLA) or experience a noticeable increase in end-to-end email delay, thereby negatively impacting the user experience. To capture the overall system's end-to-end delay, we introduce a frontend node denoted as $X_f$, which encompasses the time interval from when an email is sent to when it is received. This particular latency metric is closely monitored by engineers, as it serves as an indicator of the system's reliability performance.

Once anomalous events are detected, specifically when the frontend node $X_f$ demonstrates high latency, the primary goal is to identify the microservices accountable for the anomaly and promptly mitigate it \cite{chen2023imdiffusion}. The RCA process aims to identify the root cause nodes within the trace $\mathcal{G}$ that are most likely to have a significant impact on the end-to-end email delay. By minimizing the time needed to pinpoint the root cause of an anomaly, RCA plays a crucial role in ensuring the high reliability of microservice systems.

Let $\mathbf{X} = (X_1, X_2, \cdots, X_n)$ represent the set of components in the service dependency graph $\mathcal{G}$, where each component collects the latency metric within a specific time window. The objective of the RCA process is to identify the subset of nodes $\mathbf{c}\subseteq(1,\cdots, n)$ that are most likely to be the root cause of the incident. We define the duration of the anomalous event as the alerting window, denoted as $W_a$. Additionally, latency data collected from a base window $W_b$, when the system is operating normally, will be utilized for comparative analysis for the RCA, as shown in the right of Fig.~\ref{fig:intro}.


\vspace*{-0.5em}
\section{Empirical Study on Trace Data\label{sec:study}}
In order to facilitate the comprehension of the RCA framework, it is necessary to examine the trace data more closely and characterize the latency behavior of the root cause service and healthy components in the system during an incident. Insights gained from the analysis can inform the design of the pruning policy and subsequent RCA process. We conduct a brief empirical analysis of historical incident instances in the Microsoft Exchange system, where one or more services caused problems leading to incidents.

The empirical study utilizes a trace dataset obtained from the Microsoft Exchange microservice system, specifically focusing on 12 instances of RCA referred to as incidents. In these incidents, a subset of services within the system exhibits high latency, leading to unexpected delays in end-to-end processing. Engineers have already identified the root causes for each of these RCA instances. The dataset covers a duration of $345$ days and includes trace data from over $708$ distinct microservices. These microservices are distributed across $11$ forests, resulting in a cumulative count of $8,424$ microservices. The substantial size of the dataset facilitates a comprehensive analysis of trace characteristics and the distinctive properties of the root cause service.

Building upon the methodology outlined in \cite{TraceArk}, we calculate two types of latency measurements for each component in the trace: \emph{(i)} Exclusive Latency (ExL), which indicates the running time of the component itself, and \emph{(ii)} Inclusive Latency (InL), which is the difference between the component entry and exit time. The InL metric is equal to the sum of the component's exclusive latency and the inclusive latency of its children. Both ExL and InL metrics are frequently used in trace studies \cite{huang2021tprof, liu2020unsupervised, nedelkoski2019anomaly}. In addition, we define a component as being \emph{``affected''} by the root cause service only if the root cause is a descendant of this node. This means that the node directly or indirectly invokes the root cause service and is thus affected. The remaining components that do not interact with the root cause service are considered \emph{``redundant''}.

\vspace*{-0.5em}
\subsection{Redundant Components for RCA\label{sec:stats}}
\begin{table}[t]
\centering
\caption{The statistics of trace data during incidents.\label{tab:stats}}
\vspace{-3mm}
\resizebox{1\columnwidth}{!}{
\begin{tabular}{lcccc} 
\hline
   \multicolumn{1}{l}{\textbf{Metric}}     & \multicolumn{1}{l}{\textbf{Total Node}} & \multicolumn{1}{l}{\textbf{Total Edge}} & \multicolumn{1}{l}{\textbf{Root Cause Num.}} & \multicolumn{1}{l}{\textbf{Affected Nodes}}  \\ 
\hline
Mean & 549                                      & 29,534                                    & 3                                        & 16                                             \\
Median  & 561                                      & 31,271                                    & 1                                        & 9                                              \\
\hline
\end{tabular}}
\vspace*{-1.5em}
\end{table}

We begin by computing the statistics of the trace data and the root cause components during incidents, as presented in Table~\ref{tab:stats}. Metrics are computed over all incident instance, similarly hereinafter. It is noteworthy that the scale of the microservice system is immense, with an average of over 500 components and approximately 30,000 invoked edges, making the RCA process highly complex. However, despite the large number of services, only a few components are responsible for causing the incidents, and only a small subset of components is affected by the root cause node. This implies that a significant number of components can be deemed redundant for the RCA process and hence can be pruned.

\begin{center}
\begin{tcolorbox}[colback=blue!5,
  colframe=gray!10,
  width=\boxwidth,
  arc=2mm, auto outer arc,
  boxrule=0.5pt,
  left=\innerwidth,
  right=\innerwidth,
]
\textbf{Takeaways 1}: A limited number of components are accountable for causing the incidents, and the impact of the root cause node is constrainted to a small subset of components, resulting in redundancy in the RCA process.
\end{tcolorbox}
\end{center}

\vspace*{-0.5em}
\subsection{Exclusive latency of Root Cause Services\label{sec:latency}}
\begin{table}[h]
\centering
\small
\vspace*{-1em}
\caption{Exclusive latency statistics of root cause and non-root cause components.\label{tab:exclusivelatency}}
\vspace{-3mm}
\begin{tabular}{lccccc}
\hline
\textbf{Service}            & \textbf{Metric} & \textbf{ExL [s]} & \textbf{Percentile}
\\ \hline
\multirow{2}{*}{Root Cause} & Mean            & 50.798

                 & P96                                        \\
                            & Median          & 1.543
                 & P96                &                                                                   \\ \hline
\multirow{2}{*}{Non-root cause}   & Mean            & 16.002             & \textbackslash{}                                              \\
                            & Median                         & 0.053 & \textbackslash{} 
                                                 \\ \hline
\end{tabular}
\vspace{-1.2em}
\end{table}

Furthermore, we perform a comparison of the exclusive latency between the root cause services and non-root cause components, as shown in Table~\ref{tab:exclusivelatency}. The objective of this analysis is to determine if ExL can effectively distinguish between these two types of components. By using ExL instead of InL, we eliminate the contribution of latency introduced by other services invoked by the target component \cite{TraceArk}. Observe that the root cause components exhibit an average ExL that is more than three times higher than that of non-root cause components. Moreover, during an incident, the ExL of the root cause components exceeds that of over 96\% of all components on average. This significant difference in ExL values establishes it as a robust indicator for identifying the true root cause. We can then utilize ExL to design additional pruning actions aimed at enhancing the efficiency of the RCA process, as will be elaborated in Sec.~\ref{sec:latency-pruning}.

\begin{center}
\begin{tcolorbox}[colback=blue!5,
  colframe=gray!10,
  width=\boxwidth,
  arc=2mm, auto outer arc,
  boxrule=0.5pt,
  left=\innerwidth,
  right=\innerwidth,
]
\textbf{Takeaways 2}: The service responsible for the root cause of an incident typically shows a high exclusive latency, exceeding that of non-root-cause components.

\end{tcolorbox}
\end{center}

\vspace*{-0.5em}
\subsection{Root Cause vs. Anomaly\label{sec:anomaly}}
Next, we analyze the correlation between the anomalous components and the root cause service when incidents occur. However, it is important to note that an anomalous component does not necessarily imply that it is the root cause of the incident, as it may have been impacted by another service that it calls \cite{guo2020graph}. To identify anomalies within the microservice system at the service level, we employ a state-of-the-art anomaly detector called \traceark \cite{TraceArk}. This detector has been integrated into Microsoft Exchange and serves as a prerequisite for the functioning of our \name. \traceark provides three intermediate metrics as indicators of anomaly, namely: \emph{(i)} \textbf{NormalizeCount}, quantifying the number of occurrences of a component normalized by the total number of traces; \emph{(ii)} \textbf{OverHead}, referring to the increment in terms of the total latency of the alerting window with respect to the base window for comparison; and \emph{(iii)} \textbf{RankScore}, indicating the degree of anomaly in the temporal dimension based on the continually in the anomaly and numerical changes derived from k-sigma threshold \cite{bernieri1996line}. Higher values of these metrics indicate a greater likelihood that the component is anomalous. Additionally, \traceark provides a direct label to predict whether a component is anomalous based on the aforementioned metrics.

\begin{table}[t]
\centering
\caption{Percentiles of anomaly indicators and identification ratio by \traceark of root causes and affected components.\label{tab:anomaly}}
\vspace{-3mm}
\resizebox{1\columnwidth}{!}{
\begin{tabular}{lccccc}
\hline
\textbf{Service}            & \textbf{Metric} & \textbf{NormalizeCount} & \textbf{RankScore} & \textbf{OverHead} & \textbf{\begin{tabular}[c]{@{}c@{}}Anomaly \\ Ratio\end{tabular}} \\ \hline
\multirow{2}{*}{Root Cause} & Mean            & P85                 & P97                & P94               & \multirow{2}{*}{90.48\%}                                          \\
                            & Median          & P83                 & P99                & P99               &                                                                   \\ \hline
\multirow{2}{*}{Affected}   & Mean            & P91                 & P92                & P93               & \multirow{2}{*}{37.49\%}                                          \\
                            & Median          & P90                 & P99                & P99               &                                                                   \\ \hline
\end{tabular}}
\vspace*{-1.5em}
\end{table}
Table~\ref{tab:anomaly} shows the percentile statistics of \textit{NormalizeCount}, \textit{RankScore}, and \textit{OverHead}, and the anomaly identification ratio by \traceark, for both the root cause service and the affected components. Percentiles are computed among all components for individual incidents. Notably, both the root cause and affected services exhibit high percentile values in terms of the three anomaly indicators. This implies that these services are very likely to be identified as anomalies. The high anomaly identification ratio by \traceark further supports this observation, with over 90\% of root cause nodes being identified as anomalies, while this ratio is much lower for affected services. This is reasonable, as other services may be impacted to different degrees, and some may not exhibit strong anomaly behaviors. Based on these insights, we leverage these anomaly indicators to prune irrelevant components, as detailed in Sec.\ref{sec:ano-pruning}.

\begin{center}
\begin{tcolorbox}[colback=blue!5,
  colframe=gray!10,
  width=\boxwidth,
  arc=2mm, auto outer arc,
  boxrule=0.5pt,
  left=\innerwidth,
  right=\innerwidth,
]
\textbf{Takeaways 3}: The anomaly indicators in \traceark show that the root cause and other services impacted by it are highly likely to have high values that indicate to anomaly. The root cause service has a higher probability of being identified as an anomaly compared to other affected components.
\end{tcolorbox}
\end{center}

\subsection{Latency Correlation Analysis \label{sec:pearson}}
Finally, we conducted an analysis of the latency similarity between the end-to-end delay observed at the frontend node and the root cause service, as well as the services affected by calculating their mutual Pearson correlation coefficients. This analysis aimed to determine whether the root cause and affected services exhibited similar latency patterns in comparison to the end-to-end delay.

\begin{table}[h]
\centering
\small
\caption{The Pearson correlation coefficient statistics between InL of the fontend node and root cause/affected services.}
\label{tab:pearson}
\vspace{-3mm}
\begin{tabular}{lccccc}
\hline
\textbf{Service}            & \textbf{Metrics} & \textbf{Pearson Coefficient} & \textbf{Percentile}
\\ \hline
\multirow{2}{*}{Root Cause} & Mean            & 0.598
                 & P83                                        \\
                            & Median          & 0.626                 & P92                &                                                                   \\ \hline
\multirow{2}{*}{Affected}   & Mean            & 0.526      & P81                                                         \\
                            & Median                         & 0.519    & P88                                              \\ \hline
\end{tabular}
\vspace*{-1.5em}
\end{table}

The statistical results are presented in Table~\ref{tab:pearson}. It can be observed that, on average, both the root cause service and the affected services displayed Pearson correlation coefficients of over $0.5$ with the frontend node, indicating a strong similarity in latency patterns. Additionally, we noted that the coefficients of the root cause services, on average, exceeded those of approximately $83\%$ of the components in the entire system. While this percentile was slightly lower for the affected services ($81\%$), the Pearson correlation coefficients of both the root cause and affected services ranked highly among all components. This valuable insight serves as the basis for designing an additional pruning action that considers the latency similarity with the end-to-end delay, as described in Sec.~\ref{sec:pearson-pruning}.

Furthermore, we observed only 7.1\% of the components with the highest Pearson correlation coefficient with the end-to-end delay, as monitored at the frontend node, were indeed the root cause of the incident. This finding highlights the fact that a high correlation does not necessarily indicate causation. Consequently, we are motivated to adopt a causal approach \cite{pearl2009causality}, as opposed to correlation-based methods, for the subsequent RCA phase. This shift in approach aims to achieve a more robust and accurate RCA process. We provide a detailed explanation of this approach in Sec.~\ref{sec:causalrca}.

\begin{center}
\begin{tcolorbox}[colback=blue!5,
  colframe=gray!10,
  width=\boxwidth,
  arc=2mm, auto outer arc,
  boxrule=0.5pt,
  left=\innerwidth,
  right=\innerwidth,
]
\textbf{Takeaways 4}: The inclusive latency of both the root cause service and other affected services typically show high correlations with the overall end-to-end delay. However, the component with the highest latency correlation with the frontend node may not necessarily be the root cause of the incident.
\end{tcolorbox}
\end{center}

\vspace*{-0.5em}
\section{The Design of \name }
We overview the architecture of \name in Sec.~\ref{sec:overview}, and delve into the details of its components in the subsequent sections.

\vspace*{-0.5em}
\subsection{\name in a Nutshell\label{sec:overview}}
\begin{figure}[t]
\centering
\includegraphics[width=\columnwidth]{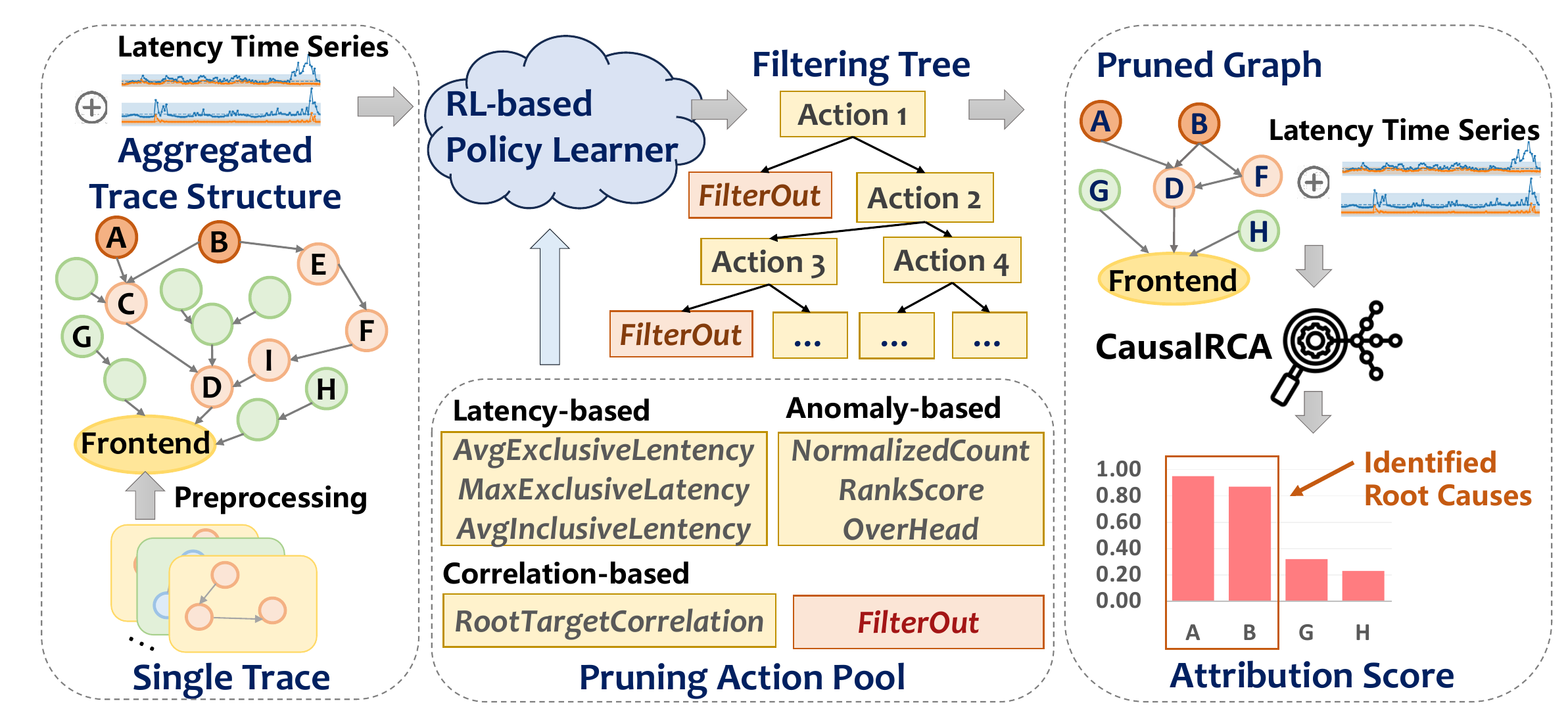}
\vspace*{-2em}
\caption{The overall framework of \name. \label{fig:framework}}
\vspace*{-2em}
\end{figure}

Fig.~\ref{fig:framework} illustrates the overall framework of \name for the RCA process applied to trace data. The Microsoft Exchange system collects a large volume of single trace data, each containing the invocation sequence and latency information of services involved in email delivery. These traces are preprocessed and aggregated to construct a dependency graph comprising hundreds of components. 

The next stage involves pruning redundant components from the dependency graph to enhance the efficiency of the subsequent RCA process. We define a library of pruning actions based on the analysis presented in Sec.~\ref{sec:study} and employ RL to automatically learn an interpretable and adaptive pruning policy from historical incidents. This learned policy is then applied to the complete service dependency graph, resulting in a simplified structure with significantly fewer components. It is important to ensure that the parent and child nodes of a pruned node are automatically connected to maintain a coherent topology structure. In Figure 1, component $E$ is pruned by the filtering tree, resulting in its parent $B$ and child $F$ being connected in the pruned graph. Further details about the pruning process can be found in Sec.~\ref{sec:pruning} and~\ref{sec:rl}.

Finally, we perform RCA using a causal approach on the pruned graph, taking into account the evolution of service latencies, to identify the root cause of the incident and attribute its responsibility. A detailed explanation of the RCA process is provided in Sec.~\ref{sec:causalrca}. In the subsequent subsection, we delve into the specifics of each component within the \name framework.

\vspace*{-0.5em}
\subsection{Trace Preprocessing\label{sec:preprocess}}
The Microsoft Exchange system generates gigabytes of trace data daily, which are collected and stored in a database for analysis. However, performing RCA on such a large amount of data is impractical and unnecessary due to the huge computing resource required, and the redundancy present in the full data. To mitigate these issues, we first randomly sample 1\% of the raw trace to maintain RCA overhead without compromising RCA performance. Note that the sampling is performed at the trace level, not at the component level. Therefore, a component is unlikely to be eliminated from the topology if it is frequently invoked. After trace sampling, we aggregate the time series for each component using the average function, normalized by the appearance of traces invokes that component at the 15-minute level. This approach offers a good tradeoff between processing overhead and RCA performance.

      


\begin{table}[t]
    \centering
    \caption{The list of pruning action pool employed in \name, and their thresholds. \label{tab:pool}}
    \vspace{-3mm}
    \small
    \begin{tabular}{c|c|c}\hline
        \textbf{Pruning Action} & \textbf{Type} & \textbf{Threshold} \\\hline
        $AvgExclusiveLentency$ & Latency & $P80$, $P85$, $P90$, $P95$, $P99$ \\
        $MaxExclusiveLentency$ & Latency & $0.01s$, $0.05s$, $0.1s$, $0.5s$, $1s$ \\
        $AvgInclusiveLentency$ & Latency & $P80$, $P85$, $P90$, $P95$, $P99$ \\
    
        $NormalizedCount$ & Anomaly & $P50$, $P65$, $P70$, $P80$, $P90$ \\
        $OverHead$ & Anomaly & $P80$, $P85$, $P90$, $P95$, $P99$ \\
        $RankScore$ & Anomaly & $P80$, $P85$, $P90$, $P95$, $P99$ \\
        $RootTargetCorrelation$ & Correlation & $0.1$, $0.3$, $0.5$, $0.7$, $0.9$ \\
        $FilterOut$ & $-$ & $-$\\\hline
    \end{tabular}
    \vspace*{-1.5em}
\end{table}

\vspace*{-0.5em}
\subsection{Service Dependency Graph Pruning\label{sec:pruning}}
Building on the insights obtained in Sec.\ref{sec:stats}, it is imperative to eliminate redundant nodes from the complete trace data as most of them are not affected by problematic services, to improve the following RCA efficiency. To this end, we propose three sets of interpretable pruning actions tailored to the trace data based on findings from Sec.~\ref{sec:study}: \emph{(i)} latency-based pruning, \emph{(ii)} anomaly-based pruning, and \emph{(iii)} correlation-based pruning. We present the complete set of pruning actions utilized in \name in Table~\ref{tab:pool}, and the subsequent subsections provide detailed explanations. These pruning actions help to remove redundant nodes while minimizing the risk of dropping true root cause and affected nodes from the service dependency graph. By doing so, the RCA efficiency can be significantly improved while maintaining high accuracy.

\subsubsection{Latency-based Pruning\label{sec:latency-pruning}}
Based on the findings discussed in Sec.~\ref{sec:latency}, where it was observed that the root cause service tends to exhibit high latency during an incident, we propose three sets of pruning actions based on this observation, as follows: \emph{(i)} \textit{AvgExclusiveLatency}, which represents the average exclusive latency of a service; \emph{(ii)} \textit{MaxExclusiveLatency}, which denotes the maximum exclusive latency of a service; and \emph{(iii)} \textit{AvgInclusiveLatency}, which indicates the average inclusive latency of a service. We include it as a pruning action for the sake of completeness. If the aforementioned metrics fall below a certain percentile or time threshold, we remove the corresponding services from the service graph, as they are unlikely to be the root cause service and are less likely to contribute to the incident.

\subsubsection{Anomaly-based Pruning\label{sec:ano-pruning}}
The second set of pruning actions considered in \name is based on the insights gained from Sec.~\ref{sec:anomaly}, which demonstrate that both the root cause service and affected services exhibit high percentile values of anomaly indicators provided by \traceark. We design three sets of pruning actions accordingly: \emph{(i)} \textit{NormalizedCount}, \emph{(ii)} \textit{RankScore}, and \emph{(iii)} \textit{OverHead}. These actions aim to filter out redundant components with low anomaly indicator percentile values compared to other components. Such components are unlikely to be the root cause or affected services, and therefore pose a low risk of being mistakenly removed. Consequently, they can be safely eliminated from the service graph without compromising the RCA performance.


\subsubsection{Correlation-based Pruning\label{sec:pearson-pruning}}
Lastly, based on the insights obtained from Sec.~\ref{sec:pearson}, we have observed that there is generally a high correlation between the end-to-end delay and both the root cause service and the affected services. This indicates that they exhibit similar latency behaviors during an incident. To leverage this observation, we introduce the last pruning action, namely \textit{RootTargetCorrelation}. This action involves evaluating the Pearson correlation coefficient between the end-to-end delay and the InL of a target node. If the correlation coefficient falls below a specified threshold, we remove the target node from the service graph. This action filters out components that demonstrate dissimilar latency patterns from the frontend node, implying that they are unlikely to be relevant to the incident.

Overall, based on the insights obtained from Sec.~\ref{sec:study}, we have developed three distinct sets of $35$ interpretable pruning actions to eliminate redundant components from the service graph, considering various characteristics of the root cause in the trace data. Additionally, we have introduced a general action, called $FilterOut$, which is responsible for executing the pruning actions associated with each filter in the filtering tree. The details of the pruning tree construction is elaborated in the subsequent Sec.~\ref{sec:rl}.

\vspace*{-0.5em}
\subsection{Learning Pruning Policy with RL\label{sec:rl}}

Once we have a comprehensive library of candidate pruning actions in Table~\ref{tab:pool}, the next step is to select appropriate actions from this pool and determine their execution sequence to create a pruning policy. The objective is to maximize the effectiveness of the RCA while minimizing the number of components in the service graph, thereby improving the efficiency of the RCA process. However, designing an optimal pruning policy is not a straightforward task.
Exploring all possible combinations of $N$ actions would require a complexity of $O(2^N)$ trials. Additionally, determining the optimal execution sequence of these actions further complicates the problem. Given the diverse nature of incidents and trace data, relying solely on human expertise to design an adaptive graph pruning policy that can generalize to unseen RCA instances is infeasible, even with a candidate list of pruning actions at hand.

To overcome this challenge, we propose a solution that utilizes a RL approach to automatically learn a pruning policy, known as a filtering tree, from historical data. This learned filtering tree can then be applied to new incidents without the need for human intervention. The RL approach offers several distinct advantages:
\begin{itemize}[leftmargin=*]
    \item \textbf{Automatic Learning}: The RL approach can automatically to learn the pruning policy without requiring human intervention. This automation reduces the reliance on manual expertise and allows for efficient policy learning. 
    \item \textbf{Effective Searching}: The RL approach naturally handles delayed rewards by employing T-D learning, enabling efficient RCA performance. Furthermore, by balancing exploration and exploitation, RL effectively explores the complex action space, resulting in significantly improved efficiency compared to exhaustive search methods.
    \item \textbf{Adaptability}: While the RL policy is represented via a parametric model, the learned filtering tree is adaptive and can be applied to new incidents without the need for further refinement by humans. This adaptability enhances the generalizability of the pruning policy to unseen scenarios.
\end{itemize}
We begin by introducing the concept of the filtering tree and subsequently provide a detailed description of our RL approaches in the following subsections.

\subsubsection{Filtering Tree}
\begin{figure}[t]
\centering
\includegraphics[width=0.9\columnwidth]{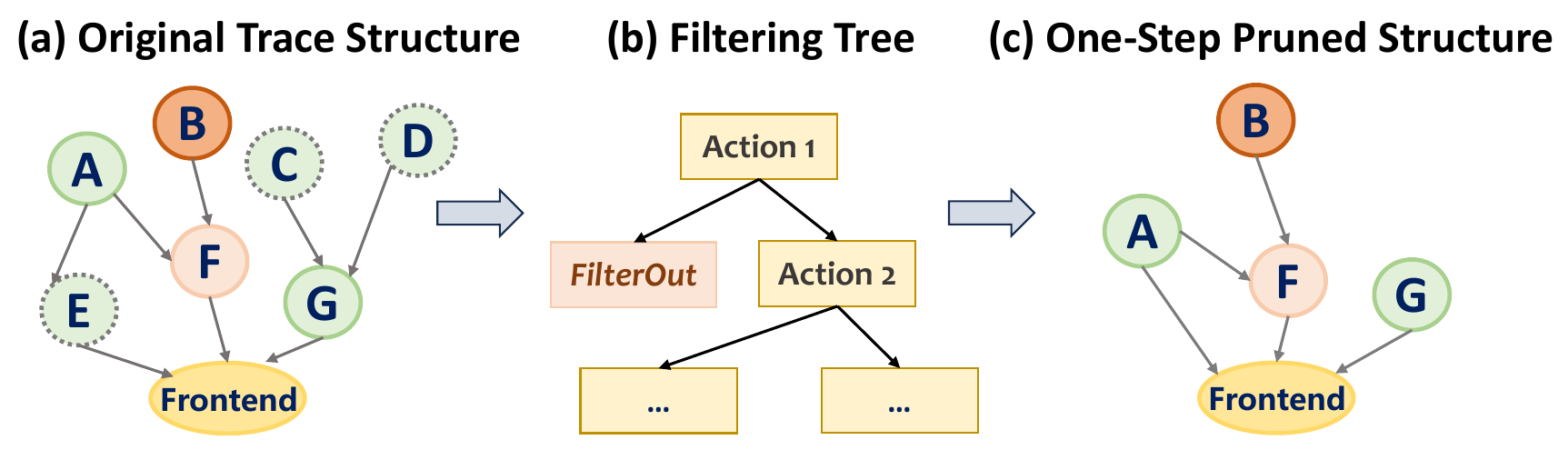}
\vspace*{-1em}
\caption{An example of a filtering tree and its execution for one-step pruning.  \textbf{C}, \textbf{D} and \textbf{E} are pruned by Action 1.
\label{fig:example}}
\vspace*{-1.5em}
\end{figure}
A filtering tree is a hierarchical structure utilized to eliminate redundant components in the service dependency graph. It is structured as a binary tree, with each node representing a pruning action selected from Table~\ref{tab:pool}. If a node has children, it divides the full component set in the dependency graph into subsets based on whether they satisfy the action's condition. Specifically, if a component surpasses the pruning action's threshold, it is directed to the right child node; otherwise, it is directed to the left child node. The \textit{FilterOut} action is exclusively present in the leaf nodes of the filtering tree, where it executes the pruning action on the components traversing its branch.

In Fig.~\ref{fig:framework}, we provide an example of a one-step pruning. The original trace consists of 7 components, excluding the frontend node. Components $C$, $D$, and $E$ are assigned to the left branch of Action 1 and meet the \textit{FilterOut} condition, leading to their removal. The remaining components that meet the threshold of Action 1 are directed to its right branch and subsequently evaluated by Action 2. By applying this filtering process to all service components in the dependency graph, redundant nodes are eliminated, resulting in a more efficient RCA procedure.


\vspace*{-1em}
\begin{figure}[!t]
\centering
\includegraphics[width=\columnwidth]{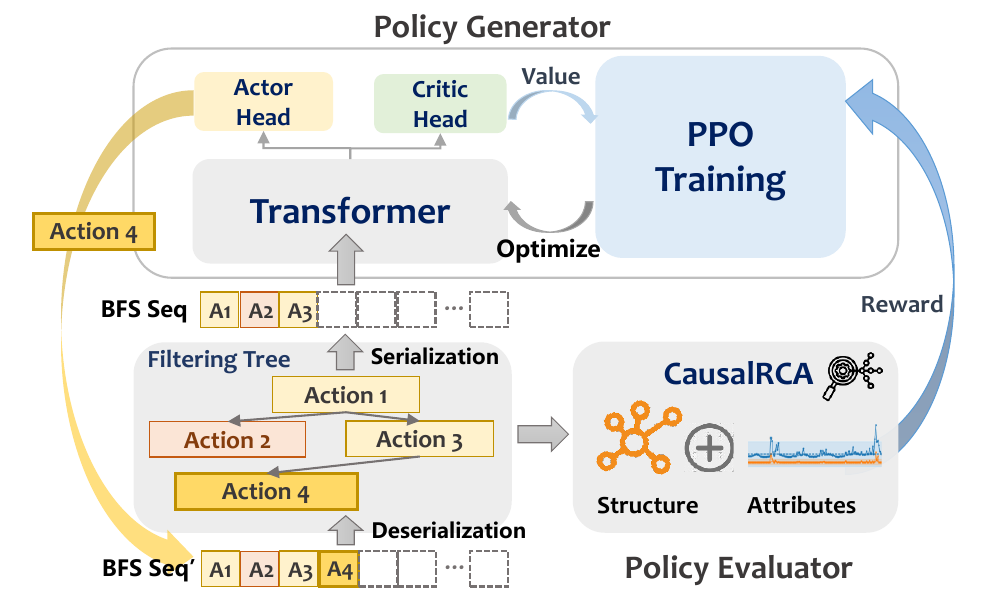}
\vspace*{-1.5em}
\caption{The RL-framework for the filtering tree training. 
\label{fig:model}}
\vspace*{-1.5em}
\end{figure}

\subsubsection{Model Overview}
The overall RL-framework employed for the construction of the filtering tree is illustrated in Fig.~\ref{fig:model}. This framework consists of two main components: \emph{(i)} a \textbf{Policy Generator}, responsible for learning a filtering tree that makes decisions regarding the selection of pruning actions and the branches to follow in order to create an effective filter; and \emph{(ii)} a \textbf{Policy Evaluator}, which uses the RCA system to evaluate the performance of the generated filtering tree. The output of the policy evaluator serves as a reward signal, which is utilized to enhance the policy generator using Proximal Policy Optimization (PPO) techniques. Following the training process, the policy generator is capable of constructing a filtering tree that achieves both high RCA accuracy and efficiency.

\subsubsection{MDP Formulation}
We start by formulating the generation of the filtering tree as a finite-horizon Markov Decision Process (MDP). The environment comprises an action space $\mathcal{A}$, a state space $\mathcal{S}$, and a reward function $R$. The filtering tree $f$ can be serialized as a sequence of nodes by visiting each node in a breadth-first manner, denoted as $\tau$. This serialization enables us to represent the filtering tree as a vector and recover it through deserialization. Therefore, we define the action search space as a discrete sequence of ``nodes'' that correspond to pruning actions selected from Table~\ref{tab:pool}.

At each time step $t$, given the current state $s_t$ of the filtering tree and the pruned trace structure, the policy generator $\pi_\theta(a_t|s_t)$ selects a pruning action $a_t$ to partition the set of components from the dependency graph. This action is considered as adding a new node to the filtering tree. Upon receiving a reward signal $r_t$ from the policy evaluator at each step, the environment transitions to a new state, resulting in a new filtering tree and pruned dependency graph. The objective of the agent is to learn the optimal policy that maximizes the expected cumulative reward by selecting the most appropriate pruning action at each step. The different components of the MDP are defined as follows:
\begin{itemize}[leftmargin=*]
    \item \textbf{State} $s_t\in\mathcal{S}$ encompasses two components: the structure of the filtering tree and the current structure, both serialized using the breadth-first search (BFS) algorithm. The complexity of the trace, which includes the number of nodes, edges, and sparsity, is also included as a metric in the state tensor.
    \item \textbf{Action} $a_t\in\mathcal{A}$ corresponds to selecting a single action from the set of actions defined in Table~\ref{tab:pool} to be added to the filtering tree.
    \item \textbf{Reward Function} $R(s_t,a_t) = \alpha \cdot r^{com} + \beta \cdot r^{rca}$ comprises two components. $r^{com} = -(|N| + |E| + \frac{|E|}{|N|^2})$, quantifies the current complexity of the pruned trace structure. Here, $\alpha$ and $\beta$ denote the weights assigned to the rewards. To be Specific, we set $\alpha = 0.01$ and $\beta = 1$. $|N|$ and $|E|$ represent the number of nodes and edges, respectively.  $r^{rca} = PR@Avg + RankScore$, measures the accuracy of the RCA based on the pruned trace structure. For a detailed explanation of $PR@Avg$ and $RankScore$, refer to Sec.~\ref{sec:metric}.
\end{itemize}
The reward function incorporates two terms, aiming to maximize RCA accuracy while minimizing graph complexity. This ensures high RCA performance while achieving efficiency improvements.

\subsubsection{Filtering Tree Generator}
Given the current state $s_t$ of the filtering tree and the pruned trace structure, the filtering tree generator $\pi_\theta(a_t|s_t)$ learns to choose the next pruning action added to the filtering tree. This action is selected from the Actor head of the model in Fig.~\ref{fig:model}. Since the filtering tree is serialized as a sequence vector, we leverage the Transformer~\cite{vaswani2017attention} as the model of $\pi_\theta$ to map the state $s_t$ into action $a_t$.

\noindent\textbf{Cascade Policy Structure:} In order to effectively remove redundant components from the trace, it is essential to prioritize the \textit{FilterOut} action. Otherwise, the resulting pruning tree would become excessively large. To address this, we introduce a cascade policy~\cite{wang2022nenya} that promotes the selection of \textit{FilterOut} actions:

\begin{align}
      \pi_\theta(a_t|s_t)= \left\{ \begin{array}{ll}
        \pi_\theta(a^{FO}_t=1|s_t) ,&\textrm{$a^{FO}_t=1$}\\
 \pi_\theta(a^{FO}_t=0|s_t)\cdot \pi_\theta(a^{FE}_t|s_t,a^{FO}_t=0),&\textrm{$a^{FO}_t=0$}
  \end{array} \right.
\end{align}
where $a^{FO}_t=1$ represents the decision of the policy generator to take the \textit{FilterOut} action, while $a^{FO}_t=0$ implies the omission of this action. On the other hand, $a^{FE}_t$ denotes the policy generator's selection of a non-\textit{FilterOut} pruning action. It is important to note that the \emph{FilterOut} action leads to an early termination of the expansion process for the filtering tree in that particular branch.

\subsubsection{In Situ Constraints}
By formulating the tree construction problem as a sequential decision-making problem, we can effectively incorporate domain knowledge by imposing constraints directly on the search space. This allows us to define a broad range of in situ constraints, where we can determine which actions are disallowed during the traversal. To achieve this, we simply assign zero probability to any pruning action that would violate a constraint before selecting the action. This approach ensures that all generated filtering trees adhere to the specified constraints during the learning process, reducing complexity and eliminating the need for post hoc rejection of samples.

To enhance the interpretability of the resulting filtering trees, we impose several constraints:
\emph{(i)} We limit the length of each filtering tree expression to a predefined range between a minimum and maximum length, such as 2 to 30.
\emph{(ii)} We prohibit the placement of the \textit{FilterOut} action as the right child of any nodes.
\emph{(iii)} We disallow a child node from being the same action as its parent to avoid redundant execution.
\emph{(iv)} To prevent invalid policies, we ensure that each expression includes at least one \textit{FilterOut} action within the latest five steps, with a probability of $0.8$. This probability gradually decays in subsequent learning episodes, following a decay rate of $0.8^{episodes + 1}$. Furthermore, to promote diversity in the expression of the filtering tree, we introduce randomness in the first $5$ episodes by randomly selecting actions with an equal probability.

\subsubsection{Policy Evaluator} 
Once we obtain a sequence $\tau$ generated by the Policy Generator, we deserialize the corresponding filtering tree structure $f$. This filtering tree is then directly applied as the modular component for pruning the trace structure. To assess the effectiveness of the policy, we execute it on multiple incident instances and calculate the average episodic reward, which includes both the RCA accuracy and the complexity of the pruned trace structure. This computed reward serves as the signal for training the Policy Generator using RL. The cumulative reward function of the Policy Generator is defined as follows:
\begin{align}
    R(\tau) = \frac{1}{N}\sum_{i=1}^N\sum_{t=1}^T \alpha \cdot r^{com}_{t,i} +\beta \cdot r^{rca}_{i}.
\end{align}

    \subsubsection{Actor Critic Algorithm Training with PPO} The reward function derived from the Policy Evaluator is non-differentiable due to its origin from a control environment and the combination of multiple objectives. Therefore, we utilize RL to optimize the Policy Generator. In such cases, auto-regressive models with black-box reward functions often employ the Proximal Policy Optimization (PPO) algorithm \cite{schulman2017proximal}, which optimizes the objective given by:
\begin{align}
    J(\theta) = E_{(s,a)\sim D_{\pi_{\theta}}}\left [\frac{\pi_\theta(a|s)}{\pi_{old}(a|s)}Q_\omega(s,a)\right ] - \lambda KL\left [\pi_{old}|\pi_\theta\right ].
\end{align}
Here, $\pi_{\text{old}}$ represents the policy model from previous training iterations, $Q_\omega$ denotes the Critic, which is also represented by a Transformer model with parameter $\omega$. It is important to note that the optimization of $J(\theta)$ aims to maximize the expected reward under the distribution. 

The training process is fully automated and does not require human intervention. After the training phase, we obtain a filtering tree that effectively removes redundant components for RCA. The trained filtering tree can generalize to unseen incidents, providing efficient RCA capabilities without the need for manual intervention.

\vspace*{-0.5em}
\subsection{Causal Root Cause Analysis\label{sec:causalrca}}
After obtaining the pruned dependency graph, we utilize it as a graphical causal model to perform RCA based on the changes in the causal mechanism of each node in case of a failure in the pruned dependency graph. This method, inspired by the research in \cite{budhathoki2021did}, is called \causalrca. The reason we adopt a causal method instead of correlation-based approaches is that a node that has the highest correlation with the end-to-end delay may not necessarily be the true root cause of the incident, as discussed in Sec.~\ref{sec:pearson}. This is because a component's latency is affected by the children services or functions it invokes, which may act as confounding factors. The causal method can minimize the impact of such effects and thus deliver more robust and accurate RCA performance.


We model the pruned service dependency graph using a graphical causal model. Bayesian models are employed to represent the latency of individual components in both the base window ($W_b$) and the alerting window ($W_a$) \cite{pearl1998graphical}. The joint distributions of all components are denoted as $\mathbf{P_X^b}$ and $\mathbf{P_X^a}$. The Bayesian models factorize into the product of marginal distributions for each component, $\mathbf{P_X} = \prod_{j=1}^n P_{X_j|PA_j}$. Here, $P_{X_j|PA_i}$ represents the causal mechanism between component $X_j$ and its immediate parent variables $PA_j$, which correspond to the services invoked by $X_j$ \cite{pearl2000models}.

Root cause nodes typically display a noticeable increase in latency in the alerting window, which in turn affects their parent services. This results in a shift in the causal mechanisms of these root cause nodes. Upon the occurrence of an incident, the joint distribution $\mathbf{P_X^a}$ transforms by replacing the original causal mechanisms of the root cause subset $\mathbf{c}$ with new ones, \ie
\begin{align}
\mathbf{P_X} = \prod _{j\notin \mathbf{c}} P_{X_j|PA_j}^a \prod _{j\in \mathbf{c}} P_{X_j|PA_j}^b,
\end{align}
where $P_{X_j|PA_j}^b$ is the new causal mechanism of node $j$ in the alerting window. In practice, both $P^a_{X_j|PA_j}$ and $P^b_{X_j|PA_j}$ are approximated with machine learning models from observational data available in the base window and detection window, respectively.

To evaluate the influence of each component on the frontend latency distribution $P_{X_f}$, we use the Shapley value \cite{shapley1953value} to measure the change in $P_{X_f}$ when an evaluation function $\Phi$ is altered. We compare the latency distributions of the frontend node in the base window $P^b_{X_f}$ and detection window $P^a_{X_f}$ using the evaluation function $\Phi$, such that $\Delta \Phi = \Phi(P^a_{X_f}) - \Phi(P^b_{X_f})$. In this work, we employ the median function as $\Phi$ to quantify the disparity between the latency distributions of the frontend node in the two windows. However, other suitable evaluation functions can be utilized.

\vspace*{-0.5em}
\section{Evaluation}
We present a thorough evaluation of \name, utilizing real production data, and comparing its performance with several state-of-the-art baselines. This section aims to address the following research questions (RQs):
\begin{itemize}[leftmargin=*]
\item \textbf{RQ1:} How effective is \name in diagnosing the root cause of incidents occurring in large-scale microservice systems?
\item \textbf{RQ2:} How does the RL-based pruning policy compare to other pruning baselines in terms of performance?
\item \textbf{RQ3:} How efficient is the \name in the production environment, and what is its execution time at each step?
\end{itemize}
We provide answers to these questions in the following subsections.

\vspace*{-0.5em}
\subsection{Experiment Setup}
We conduct experiments using the \texttt{PyTorch} \cite{paszke2019pytorch} and \texttt{Dowhy} \cite{dowhypaper, dowhy_gcm} frameworks. The experiments are performed on a system running Ubuntu $22.04.1$, equipped with an Intel(R) Xeon(R) Gold $6338$ CPU consisting of 127 cores with a clock speed of $2.00$GHz.
\subsubsection{Dataset}
For evaluating the performance of \name, we employ a real dataset obtained from the Microsoft Exchange microservice system, as described in Sec.~\ref{sec:study} of this paper. Due to the unavailability of large-scale trace data specifically designed for RCA in the public domain, we rely on this dataset. The dataset encompasses various incident cases, out of which we randomly select $3$ cases for training the RL-based pruning policy. The remaining $9$ cases are kept separate for the purpose of testing. It is important to emphasize that all procedures involved in data collection strictly adhered to relevant regulations, ensuring the preservation of complete anonymity and desensitization of the datasets. These measures effectively address any potential privacy concerns associated with the utilization of the dataset in this study.

\subsubsection{Baselines}
We conducted a comparative analysis of the \name with a broad range of baselines. For the graph pruning phase, exhausted search or grid search is not applicable due to the huge action space. We chose two alternative methods for comparison: random pruning and a heuristic pruning strategy designed by experienced engineers, outlined below:
\begin{itemize}[leftmargin=*]
    \item \textbf{Random:} This method randomly selects pruning actions from the action pool in Table~\ref{tab:pool} to filter out components that do not meet the requirement. The search continues until the number of nodes in the pruned graph is below a threshold.
    \item \textbf{Heuristic~\cite{liu2021microhecl}:} The Heuristic method is designed by engineers who have years of experience in the field of RCA. We leverage their expertise to design the heuristic pruning rules based on their domain knowledge.
\end{itemize}

In terms of the RCA phase, we selected state-of-the-art trace-based and monitoring-based RCA approaches, which leverage correlations and graph theory, as listed below:
\begin{itemize}[leftmargin=*]
    \item \textbf{BackTrace~\cite{pan2021faster}}: This method employs a backward BFS to evaluate abnormality scores. It takes into account two key factors: the strength of path correlation originating from the frontend service and the Pearson correlation coefficient between individual components and the frontend service.
    \item \textbf{GrootRank~\cite{wang2021groot}}: It is a customized algorithm from PageRank~\cite{sanderson2010christopher} to calculate the root cause ranking. It incorporates access distances from the frontend to handle situations where tied results may occur. In this ranking, higher priority is given to root causes with shorter access distances (sum) and vice versa.
\end{itemize}

\subsubsection{Performance Metrics\label{sec:metric}}
In order to evaluate the performance of our method, we utilized two metrics to assess the precision of the RCA results: \textbf{PR@k}\cite{wang2018cloudranger} and \textbf{RankScore}\cite{pan2021faster}.
\textbf{PR@k} measures the number of correct root causes among the top-k predictions and is calculated as follows:
\begin{equation}
\text{PR@k}=\frac{\sum_{i=1}^k \mathrm{RC}_i^{\text {pred }} \in \mathrm{RC}^{\text {true }}}{\min \left(\left|\mathrm{RC}^{\text {true }}\right|, k\right)},
\end{equation}
where $\mathrm{RC}^{\text {true }}$ represents the ground-truth list of root causes and $\mathrm{RC}^{\text {pred }}$ is the set of predicted root causes. Additionally, we compute the average of \textbf{PR@k} for $k=1, 2, \cdots ,5$ as \textbf{PR@Avg}.
Another metric used to assess the rankings of the RCA is \textbf{RankScore}, defined as $=\frac{1}{\left|\mathrm{RC}^{\text {true }}\right|} \sum_{v \in \mathrm{RC}^{\text {true }}} (1-\mathrm{s}\left(v\right))$. Here, $\mathrm{s}\left(v\right)$ represents $\frac{\operatorname{rank}(v)}{N}$ when $\operatorname{rank}(v) <$ the number of true root causes. Otherwise, $\mathrm{s}\left(v\right)$ is assigned as $1$. In this context, $N$ denotes the length of the predicted root cause list, and $\operatorname{rank}(v)$ signifies the rank of $v$. The RankScore metric averages the rank performance for each ground-truth root cause service.
Furthermore, to assess the recall quality of the pruned graph, we introduce the \textbf{HitRootCause} metric, defined as $= \frac{|RC^{\text{hit}}|}{|RC^{\text{true}}|}$, where $|RC^{\text{hit}}|$ denotes the number of ground-truth root causes appearing in the pruned graph, and $|RC^{\text{true}}|$ represents the total number of ground-truth root causes. A lower value of HitRootCause indicates that the root cause services are not preserved in the pruned graph, rendering the subsequent RCA  infeasible.


\vspace*{-0.5em}
\subsection{RCA Performance Comparison\label{sec:rca}}
Table~\ref{table:rca_results} presents the comprehensive performance evaluation of our proposed \name approach compared to other baselines using 9 testing cases. All evaluations are performed on the pruned dependency graph obtained from our RL-based approach. Our findings demonstrate the superiority of \name, which leverages \causalrca method. It consistently outperforms the other baselines across all performance metrics. Specifically, \name achieves a significant improvement of at least $72.3\%$ in \textbf{PR@Avg} and $20.6\%$ in \textbf{RankScore}, indicating its enhanced ability to accurately rank the true root cause higher. A closer examination of the performance reveals that \name successfully identifies the true root cause as the top-ranked recommendation in $66.7\%$ of cases. Furthermore, its top-5 recommendations encompass $95.6\%$ of the true root causes, which is a remarkable achievement. 

These notable improvements highlight the superiority of the \causalrca approach, showcasing its robustness and accuracy in RCA. In contrast, the correlation-based approach, such as BackTrace, falls short in capturing the complex relationship between service latency and its root cause. Correlation alone is insufficient to explain the latency patterns observed across different services. The presence of confounding factors that contribute to high latency in multiple services can lead to misleading conclusions when relying solely on correlation-based RCA methods. Hence, our findings emphasize the reliability and effectiveness of causal approaches in RCA, aligning with the insights gained from Sec.~\ref{sec:pearson}.

\begin{table}
\centering
\caption{Comparison of different RCA approaches. \label{table:rca_results}}
\vspace{-2mm}
\small
\resizebox{1\columnwidth}{!}{
\begin{tblr}{
  cell{2}{2} = {c},
  cell{2}{3} = {c},
  cell{2}{4} = {c},
  cell{2}{5} = {c},
  cell{2}{6} = {c},
  cell{3}{2} = {c},
  cell{3}{3} = {c},
  cell{3}{4} = {c},
  cell{3}{5} = {c},
  cell{3}{6} = {c},
  cell{4}{2} = {c},
  cell{4}{3} = {c},
  cell{4}{4} = {c},
  cell{4}{5} = {c},
  cell{4}{6} = {c},
  hline{1-2,5} = {-}{},
}
\textbf{RCA Method} & \textbf{PR@1}    & \textbf{PR@3}   & \textbf{PR@5}  & \textbf{PR@Avg} & \textbf{RankScore} \\
BackTrace           & 0.222                  & 0.481                   & 0.689          & 0.484               & 0.678              \\
GrootRank            & 0.111                   & 0.259                   & 0.578          & 0.306               & 0.492              \\
\name              & \textbf{0.667} & \textbf{0.852}  & \textbf{0.956} & \textbf{0.834}      & \textbf{0.818}     
\end{tblr}}
\vspace*{-2.5em}
\end{table}

\vspace*{-0.5em}
\subsection{Dependency Graph Pruning Comparison}
Table~\ref{table:pruning_results} provides a comparison of the impact and performance of different graph pruning methods, namely Random, Heuristic, and \name (RL-based), across $9$ testing cases. The RL-based approach demonstrates remarkable convergence efficiency, converging in only $36$ episodes despite being trained on only three incidents. On average, \name retains only 11 components after pruning, which is equivalent to the number of components retained by the Heuristic approach designed by experienced engineers. This represents a significant reduction ($98\%$) compared to the original full dependency graph, which initially consists of over 500 components.

Despite the substantial reduction, the pruned graph retains $92.9\%$ of the root cause services (\textbf{HitRootCause}). This indicates that the majority of pruned components are indeed redundant and do not contribute to the RCA process. In contrast, the Heuristic method retains only $83.3\%$ of the root cause services in the final pruned graph. Consequently, performing RCA on the remaining one-sixth of incidents would not yield the root cause, as those components have already been removed from the data.

The impact of the pruning method is also reflected in the subsequent RCA performance. Our \name approach achieves a $26.5\%$ improvement in \textbf{PR@5}, a 26.3\% improvement in \textbf{PR@Avg}, and a $33.4\%$ improvement in \textbf{RankScore} compared to the Heuristic pruning method. Notably, these improvements surpass those achieved by the Random pruning method in an larger margin. These results demonstrate the superior performance of the RL-based pruning approach in accurately removing redundant components, leading to significant enhancements in the efficiency and effectiveness of the RCA process. Notably, an important advantage of \name is its ability to generalize well to unseen cases, as it can be trained on only three incidents and still achieve satisfactory performance on the nine testing cases. Moreover, \name converges rapidly in a small number of trials (episodes), making the pruning process fully automated, efficient, and adaptable to the RCA task.

\begin{figure}[t]
\centering
\includegraphics[width=0.9\columnwidth]{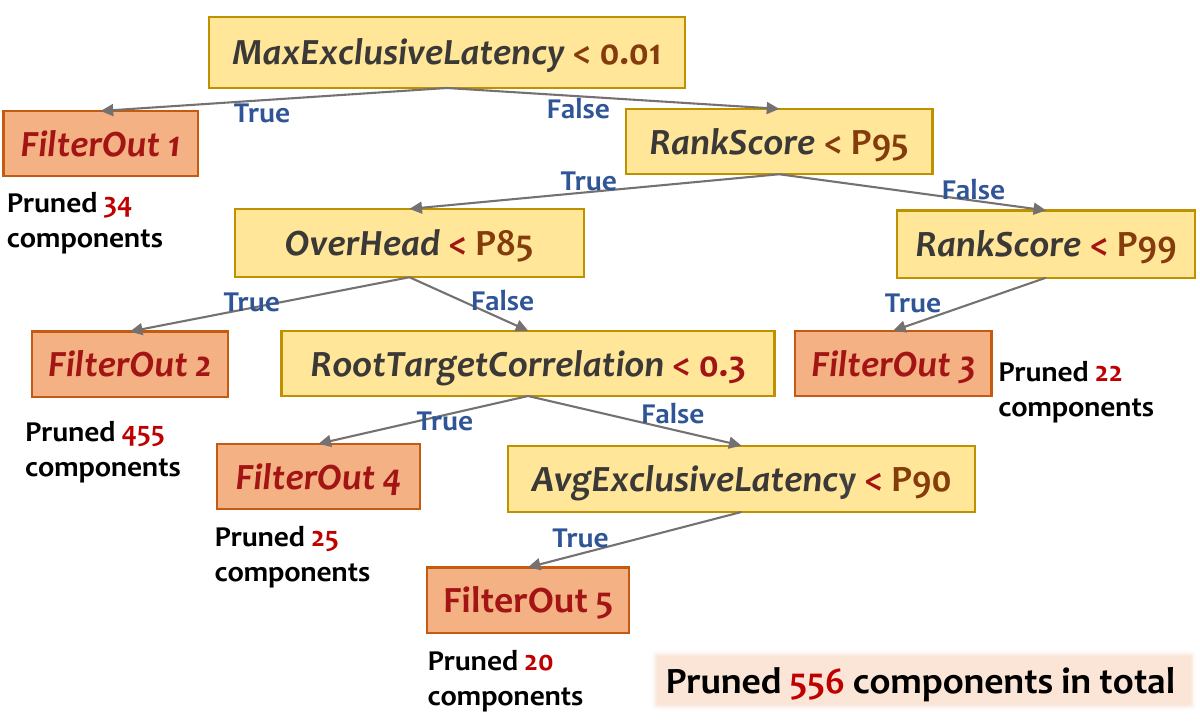}
\vspace*{-1em}
\caption{The filtering tree learned by our \name \label{fig:filtering_tree} with averaged pruned components by each FilterOut action.\label{table:efficiency}}
\vspace*{-0.5em}
\end{figure}

Finally, we conduct a detailed analysis of the filtering tree learned by our \name to gain insights into the pruning policy acquired through the RL approach. Fig.~\ref{fig:filtering_tree} illustrates the learned filtering tree, which provides a clear interpretation of the pruning policy. The average number of pruned components across all testing cases for each \textit{FilterOut} action is also show in the figure.  The filtering tree consists of a total of 11 pruning actions, with 5 of them being \textit{FilterOut} actions. It is evident that all \textit{FilterOut} actions contribute to the final pruning results by removing redundant components from the dependency graph to varying extents. Furthermore, the pruning actions in the filtering tree encompass all three types of actions specified in the action library (Table~\ref{tab:pool}), indicating their relevance and interpretability in the pruning process. However, we observe that the second \textit{FilterOut} action, which combines the joint filter actions of \textit{RankScore} and \textit{Overhead} derived from \traceark, removes a significant majority of components on average (81.8\%). This finding underscores the pivotal role of anomaly-based pruning, emphasizing the importance of the insights gained in Sec.~\ref{sec:anomaly}.


\begin{table}
\centering
\caption{Comparison of different graph pruning approaches. \label{table:pruning_results}}
\vspace{-2mm}
\resizebox{1\columnwidth}{!}{
\begin{tblr}{
  cell{2}{2} = {c},
  cell{2}{3} = {c},
  cell{2}{4} = {c},
  cell{2}{5} = {c},
  cell{2}{6} = {c},
  cell{3}{2} = {c},
  cell{3}{3} = {c},
  cell{3}{4} = {c},
  cell{3}{5} = {c},
  cell{3}{6} = {c},
  cell{4}{2} = {c},
  cell{4}{3} = {c},
  cell{4}{4} = {c},
  cell{4}{5} = {c},
  cell{4}{6} = {c},
  hline{1-2,5} = {-}{},
}
\textbf{Method} & \textbf{Node} & \textbf{HitRootCause}   & \textbf{PR@5}  & \textbf{PR@Avg} & \textbf{RankScore} \\
Random          & 20            & 0.500          & 0.289          & 0.300           & 0.355              \\
Heuristic       & \textbf{11}   & 0.833          & 0.756          & 0.660           & 0.613              \\
\name              & \textbf{11}   & \textbf{0.929} & \textbf{0.956} & \textbf{0.834}  & \textbf{0.818}     
\end{tblr}}
\vspace*{-1.5em}
\end{table}

\vspace*{-0.5em}
\subsection{Efficiency Evaluation}
Furthermore, \name exhibits high efficiency as an RCA framework. Once the filtering tree is obtained through the RL approach, the execution of the filtering tree on the original trace data has an average runtime of only 1.43 seconds. Subsequently, the subsequent \causalrca process requires an average of 38.24 seconds, resulting in a total runtime of 39.67 seconds for \name. This indicates that \name can achieve excellent RCA performance in large-scale microservice systems with remarkable efficiency. Consequently, it significantly reduces the efforts required by engineers for conducting RCA tasks.


\vspace*{-0.5em}
\section{Deployment \& Discussion}
We present the production impact of \name and discuss relevant threats to validity in this section.
\begin{table}[t]
\caption{Performance of \name in real productions. \label{tab:online_performance}}
\vspace{-3mm}
\begin{tabular}{lccc}
\hline
\multirow{2}{*}{\textbf{Framework}} & \multirow{2}{*}{\textbf{PR@Avg}} & \multicolumn{2}{c}{\textbf{Diagnosis time reduction}} \\
                                    &                                  & Easy cases               & Hard cases                 \\ \hline
Heuristic                           & 0.469                            & \multirow{2}{*}{50\%}    & \multirow{2}{*}{96.5\%}    \\
\name                & 0.821                            &                          &                            \\ \hline
\end{tabular}
\vspace*{-1.em}
\end{table}

\vspace*{-0.5em}
\subsection{Production Impact}
The proposed framework, \name has been integrated as a crucial component within the Microsoft Exchange system to facilitate the localization of root causes in service incidents. Whenever a regression in the end-to-end latency SLA is detected, indicating the occurrence of an incident, \name is activated and initiates the workflow depicted in Fig.~\ref{fig:framework}. To facilitate the extraction of contrastive patterns, we use one-day data from the incident and utilize the preceding seven days' data as the base window.

Table~\ref{tab:online_performance} presents the online performance of \name in a production environment. Due to confidentiality policies within the company, the specific number of incidents is not disclosed. In comparison to the heuristic RCA approach employed online, the new \name framework achieves a noteworthy improvement of 75.1\% in \textbf{PR@Avg}. In incidents where the root cause is easily identifiable, \name significantly reduces the required RCA time by 50\%. For incidents that pose challenges in diagnosing the root cause and necessitate further investigation by engineers, \name still attains remarkable RCA accuracy while drastically reducing the diagnosis time by over 96.5\% compared to manual efforts. These outcomes underscore the exceptional performance of \name in a production setting, leading to significantly enhanced RCA accuracy and substantial reductions in the time to diagnosis.

\vspace*{-0.5em}
\subsection{Threats to Validity}
\noindent\textbf{Internal threat:} The internal threat primarily relates to the accuracy of root cause labels and the implementation of \name. The root cause labels in our datasets are assigned by experienced domain experts based on incident reports, ensuring their reliability. Moreover, our approach provides the top five candidate root causes, aiding domain experts in decision-making. To mitigate implementation threats, we utilize mature RL frameworks \cite{schulman2017proximal}, and the code and configurations have been thoroughly reviewed by two authors.

\noindent\textbf{External threat:} Our study and experiments are conducted using Microsoft Exchange traces. However, our approach can be easily applied to other systems that have open telemetry traces.
While the incidents from the Exchange system may represent a limited range of root causes, considering its extensive size and complexity, we believe our findings are generalizable. In the future, we plan to extend our analyses to encompass a broader range of cloud systems.

\vspace*{-0.5em}
\section{Related Work}

\noindent\textbf{RL for Structure Optimization}
Previous research has explored the use of RL in discovering and refining neural network structures and various graph scenarios. In the field of neural architecture search (NAS), RL has been employed in AutoML studies \cite{zoph2016neural,he2021automl,pham2018efficient} to optimize neural network architectures. The RL agent aims to maximize network performance in tasks like image classification or language translation while minimizing computational costs. RL has also gained popularity in graph structure optimization \cite{pu2021learning,munikoti2022challenges}. In social networks \cite{li2023network}, RL has been utilized to optimize graph structures for information diffusion \cite{meirom2021controlling}, community detection \cite{alipour2022multiagent}, and influence maximization \cite{chen2021contingency}. RL agents dynamically add or remove edges between individuals to maximize objectives such as effective information propagation and identification of cohesive communities. In contrast, our \name framework employs RL to learn a pruning policy for trace data, focusing on service dependency graphs for the purpose of RCA. 

 
\noindent\textbf{Root Cause Analysis with Trace Data} Traces are essential in diagnosing system failures in complex cloud systems \cite{ma2020diagnosing, li2021practical, liu2020unsupervised}, especially under the large-scale, modern microservice system. Trace-based RCA primarily employs two methods to construct the dependency graph: the topological graph and the causal graph \cite{li2022mining}. A topological graph represents the physical or logical service dependency graph and can be either a static graph stored in a Configuration Management Database or a dynamically constructed service call graph \cite{liu2021microhecl}. On the other hand, causal graph-based approaches use causal discovery algorithms, such as the PC algorithm \cite{kalisch2007estimating}, to identify relationships among service components. Various inference methods, such as random walk \cite{ma2020automap} or DFS \cite{chen2014causeinfer}, are then applied to identify root causes. Interested readers may refer to a recent survey \cite{guo2020survey} for further details.
Our approach \name, which is orthogonal to both topological graph and causal graph-related approaches, focuses on the root cause filtering steps.

\vspace*{-0.5em}
\section{Conclusion}
This paper introduces \name, an end-to-end RCA framework for large-scale microservice systems based on trace data. We design a library of pruning actions derived from an extensive analysis of trace data and the behavior of root cause services, and then leverages RL techniques to autonomously learn an interpretable and adaptive filtering tree that eliminates redundant components from the service dependency graph. This approach significantly enhances RCA efficiency while maintaining RCA performance.  The framework further employs a \causalrca approach, which ensures accurate and robust root cause diagnosis for incidents. Experimental evaluations conducted on real trace datasets validate the efficacy of \name, highlighting substantial improvements in RCA performance. Notably, \name has been deployed as a core RCA engine within the Microsoft Exchange microservice system, where it consistently achieves over 75\% higher RCA accuracy compared to the legacy RCA framework, and effectively reduces the time required for RCA by up to 96.5\%.


\newpage
\bibliographystyle{unsrt}
\balance 
\bibliography{sample-base}

\begin{thebibliography}{10}

\bibitem{soldani2022anomaly}
Jacopo Soldani and Antonio Brogi.
\newblock Anomaly detection and failure root cause analysis in (micro)
  service-based cloud applications: A survey.
\newblock {\em ACM Computing Surveys (CSUR)}, 55(3):1--39, 2022.

\bibitem{liu2019jcallgraph}
Haifeng Liu, Jinjun Zhang, Huasong Shan, Min Li, Yuan Chen, Xiaofeng He, and
  Xiaowei Li.
\newblock Jcallgraph: tracing microservices in very large scale container cloud
  platforms.
\newblock In {\em Cloud Computing--CLOUD 2019: 12th International Conference,
  Held as Part of the Services Conference Federation, SCF 2019, San Diego, CA,
  USA, June 25--30, 2019, Proceedings 12}, pages 287--302. Springer, 2019.

\bibitem{gan2019seer}
Yu~Gan, Yanqi Zhang, Kelvin Hu, Dailun Cheng, Yuan He, Meghna Pancholi, and
  Christina Delimitrou.
\newblock Seer: Leveraging big data to navigate the complexity of performance
  debugging in cloud microservices.
\newblock In {\em Proceedings of the twenty-fourth international conference on
  architectural support for programming languages and operating systems}, pages
  19--33, 2019.

\bibitem{musk}
Elon Mask.
\newblock There are ~1200 “microservices” server side.
\newblock \url{https://twitter.com/elonmusk/status/1592561366493442050}.

\bibitem{TraceArk}
Zhengran Zeng, Yuqun Zhang, Yong Xu, Minghua Ma, Bo~Qiao, Wentao Zou, Qingjun
  Chen, Meng Zhang, Xu~Zhang, Hongyu Zhang, Xuedong Gao, Hao Fan, Saravan
  Rajmohan, Qingwei Lin, and Dongmei Zhang.
\newblock Traceark: Towards actionable performance anomaly alerting for online
  service systems.
\newblock In {\em To appear in Proc. of ICSE}, 2023.

\bibitem{liu2021microhecl}
Dewei Liu, Chuan He, Xin Peng, Fan Lin, Chenxi Zhang, Shengfang Gong, Ziang Li,
  Jiayu Ou, and Zheshun Wu.
\newblock Microhecl: High-efficient root cause localization in large-scale
  microservice systems.
\newblock In {\em 2021 IEEE/ACM 43rd International Conference on Software
  Engineering: Software Engineering in Practice (ICSE-SEIP)}, pages 338--347.
  IEEE, 2021.

\bibitem{ma2022empirical}
Minghua Ma, Yudong Liu, Yuang Tong, Haozhe Li, Pu~Zhao, Yong Xu, Hongyu Zhang,
  Shilin He, Lu~Wang, Yingnong Dang, et~al.
\newblock An empirical investigation of missing data handling in cloud node
  failure prediction.
\newblock In {\em Proceedings of the 30th ACM Joint European Software
  Engineering Conference and Symposium on the Foundations of Software
  Engineering}, pages 1453--1464, 2022.

\bibitem{ma2018robust}
Minghua Ma, Shenglin Zhang, Dan Pei, Xin Huang, and Hongwei Dai.
\newblock Robust and rapid adaption for concept drift in software system
  anomaly detection.
\newblock In {\em 2018 IEEE 29th International Symposium on Software
  Reliability Engineering (ISSRE)}, pages 13--24. IEEE, 2018.

\bibitem{liu2022uniparser}
Yudong Liu, Xu~Zhang, Shilin He, Hongyu Zhang, Liqun Li, Yu~Kang, Yong Xu,
  Minghua Ma, Qingwei Lin, Yingnong Dang, et~al.
\newblock Uniparser: A unified log parser for heterogeneous log data.
\newblock In {\em Proceedings of the ACM Web Conference 2022}, pages
  1893--1901, 2022.

\bibitem{zhou2018fault}
Xiang Zhou, Xin Peng, Tao Xie, Jun Sun, Chao Ji, Wenhai Li, and Dan Ding.
\newblock Fault analysis and debugging of microservice systems: Industrial
  survey, benchmark system, and empirical study.
\newblock {\em IEEE Transactions on Software Engineering}, 47(2):243--260,
  2018.

\bibitem{zhang2021halo}
Xu~Zhang, Chao Du, Yifan Li, Yong Xu, Hongyu Zhang, Si~Qin, Ze~Li, Qingwei Lin,
  Yingnong Dang, Andrew Zhou, et~al.
\newblock Halo: Hierarchy-aware fault localization for cloud systems.
\newblock In {\em Proceedings of the 27th ACM SIGKDD Conference on Knowledge
  Discovery \& Data Mining}, pages 3948--3958, 2021.

\bibitem{ma2020diagnosing}
Minghua Ma, Zheng Yin, Shenglin Zhang, Sheng Wang, Christopher Zheng, Xinhao
  Jiang, Hanwen Hu, Cheng Luo, Yilin Li, Nengjun Qiu, et~al.
\newblock Diagnosing root causes of intermittent slow queries in cloud
  databases.
\newblock {\em Proceedings of the VLDB Endowment}, 13(8):1176--1189, 2020.

\bibitem{wang2023rca}
Lu~Wang, Chaoyun Zhang, Ruomeng Ding, Yong Xu, Qihang Chen, Wentao Zou, Qingjun
  Chen, Meng Zhang, Xuedong Gao, Hao Fan, Saravan Rajmohan, Qingwei Lin, and
  Dongmei Zhang.
\newblock Root cause analysis for microservice systems via hierarchical
  reinforcement learning from human feedback.
\newblock In {\em Proceedings of the 29th ACM SIGKDD Conference on Knowledge
  Discovery and Data Mining}, page 5116–5125, New York, NY, USA, 2023.
  Association for Computing Machinery.

\bibitem{wang2018cloudranger}
Ping Wang, Jingmin Xu, Meng Ma, Weilan Lin, Disheng Pan, Yuan Wang, and Pengfei
  Chen.
\newblock Cloudranger: Root cause identification for cloud native systems.
\newblock In {\em 2018 18th IEEE/ACM International Symposium on Cluster, Cloud
  and Grid Computing (CCGRID)}, pages 492--502. IEEE, 2018.

\bibitem{budhathoki2021did}
Kailash Budhathoki, Dominik Janzing, Patrick Bloebaum, and Hoiyi Ng.
\newblock Why did the distribution change?
\newblock In {\em International Conference on Artificial Intelligence and
  Statistics}, pages 1666--1674. PMLR, 2021.

\bibitem{guo2020graph}
Xiaofeng Guo, Xin Peng, Hanzhang Wang, Wanxue Li, Huai Jiang, Dan Ding, Tao
  Xie, and Liangfei Su.
\newblock Graph-based trace analysis for microservice architecture
  understanding and problem diagnosis.
\newblock In {\em Proceedings of the 28th ACM Joint Meeting on European
  Software Engineering Conference and Symposium on the Foundations of Software
  Engineering}, pages 1387--1397, 2020.

\bibitem{zhou2019latent}
Xiang Zhou, Xin Peng, Tao Xie, Jun Sun, Chao Ji, Dewei Liu, Qilin Xiang, and
  Chuan He.
\newblock Latent error prediction and fault localization for microservice
  applications by learning from system trace logs.
\newblock In {\em Proceedings of the 2019 27th ACM Joint Meeting on European
  Software Engineering Conference and Symposium on the Foundations of Software
  Engineering}, pages 683--694, 2019.

\bibitem{wang2019grano}
Hanzhang Wang, Phuong Nguyen, Jun Li, Selcuk Kopru, Gene Zhang, Sanjeev
  Katariya, and Sami Ben-Romdhane.
\newblock Grano: Interactive graph-based root cause analysis for cloud-native
  distributed data platform.
\newblock {\em Proceedings of the VLDB Endowment}, 12(12):1942--1945, 2019.

\bibitem{brandon2020graph}
{\'A}lvaro Brand{\'o}n, Marc Sol{\'e}, Alberto Hu{\'e}lamo, David Solans,
  Mar{\'\i}a~S P{\'e}rez, and Victor Munt{\'e}s-Mulero.
\newblock Graph-based root cause analysis for service-oriented and microservice
  architectures.
\newblock {\em Journal of Systems and Software}, 159:110432, 2020.

\bibitem{yu2023cmdiagnostor}
Qingyang Yu, Changhua Pei, Bowen Hao, Mingjie Li, Zeyan Li, Shenglin Zhang,
  Xianglin Lu, Rui Wang, Jiaqi Li, Zhenyu Wu, et~al.
\newblock Cmdiagnostor: An ambiguity-aware root cause localization approach
  based on call metric data.
\newblock In {\em Proceedings of the ACM Web Conference 2023}, pages
  2937--2947, 2023.

\bibitem{zhang2020microscope}
Chaoyun Zhang, Marco Fiore, Cezary Ziemlicki, and Paul Patras.
\newblock Microscope: mobile service traffic decomposition for network slicing
  as a service.
\newblock In {\em Proceedings of the 26th Annual International Conference on
  Mobile Computing and Networking}, pages 1--14, 2020.

\bibitem{zhang2017zipnet}
Chaoyun Zhang, Xi~Ouyang, and Paul Patras.
\newblock Zipnet-gan: Inferring fine-grained mobile traffic patterns via a
  generative adversarial neural network.
\newblock In {\em Proceedings of the 13th International Conference on emerging
  Networking EXperiments and Technologies}, pages 363--375, 2017.

\bibitem{zhang2021cloudlstm}
Chaoyun Zhang, Marco Fiore, Iain Murray, and Paul Patras.
\newblock Cloudlstm: A recurrent neural model for spatiotemporal point-cloud
  stream forecasting.
\newblock In {\em Proceedings of the AAAI Conference on Artificial
  Intelligence}, volume~35, pages 10851--10858, 2021.

\bibitem{zhang2019deep}
Chaoyun Zhang, Paul Patras, and Hamed Haddadi.
\newblock Deep learning in mobile and wireless networking: A survey.
\newblock {\em IEEE Communications surveys \& tutorials}, 21(3):2224--2287,
  2019.

\bibitem{yu2018spatio}
Bing Yu, Haoteng Yin, and Zhanxing Zhu.
\newblock Spatio-temporal graph convolutional networks: a deep learning
  framework for traffic forecasting.
\newblock In {\em Proceedings of the 27th International Joint Conference on
  Artificial Intelligence}, pages 3634--3640, 2018.

\bibitem{ali2022exploiting}
Ahmad Ali, Yanmin Zhu, and Muhammad Zakarya.
\newblock Exploiting dynamic spatio-temporal graph convolutional neural
  networks for citywide traffic flows prediction.
\newblock {\em Neural networks}, 145:233--247, 2022.

\bibitem{mohamed2020social}
Abduallah Mohamed, Kun Qian, Mohamed Elhoseiny, and Christian Claudel.
\newblock Social-stgcnn: A social spatio-temporal graph convolutional neural
  network for human trajectory prediction.
\newblock In {\em Proceedings of the IEEE/CVF conference on computer vision and
  pattern recognition}, pages 14424--14432, 2020.

\bibitem{pan2021faster}
Yicheng Pan, Meng Ma, Xinrui Jiang, and Ping Wang.
\newblock Faster, deeper, easier: crowdsourcing diagnosis of microservice
  kernel failure from user space.
\newblock In {\em Proceedings of the 30th ACM SIGSOFT International Symposium
  on Software Testing and Analysis}, pages 646--657, 2021.

\bibitem{meng2020localizing}
Yuan Meng, Shenglin Zhang, Yongqian Sun, Ruru Zhang, Zhilong Hu, Yiyin Zhang,
  Chenyang Jia, Zhaogang Wang, and Dan Pei.
\newblock Localizing failure root causes in a microservice through causality
  inference.
\newblock In {\em 2020 IEEE/ACM 28th International Symposium on Quality of
  Service (IWQoS)}, pages 1--10. IEEE, 2020.

\bibitem{granger1969investigating}
Clive~WJ Granger.
\newblock Investigating causal relations by econometric models and
  cross-spectral methods.
\newblock {\em Econometrica: journal of the Econometric Society}, pages
  424--438, 1969.

\bibitem{long2021air}
Chan~Li Long, Yash Guleria, and Sameer Alam.
\newblock Air passenger forecasting using neural granger causal google trend
  queries.
\newblock {\em Journal of Air Transport Management}, 95:102083, 2021.

\bibitem{wu2020connecting}
Zonghan Wu, Shirui Pan, Guodong Long, Jing Jiang, Xiaojun Chang, and Chengqi
  Zhang.
\newblock Connecting the dots: Multivariate time series forecasting with graph
  neural networks.
\newblock In {\em Proceedings of the 26th ACM SIGKDD international conference
  on knowledge discovery \& data mining}, pages 753--763, 2020.

\bibitem{chen2023imdiffusion}
Yuhang Chen, Chaoyun Zhang, Minghua Ma, Yudong Liu, Ruomeng Ding, Bowen Li,
  Shilin He, Saravan Rajmohan, Qingwei Lin, and Dongmei Zhang.
\newblock {ImDiffusion}: Imputed diffusion models for multivariate time series
  anomaly detection.
\newblock {\em arXiv preprint arXiv:2307.00754}, 2023.

\bibitem{huang2021tprof}
Lexiang Huang and Timothy Zhu.
\newblock tprof: Performance profiling via structural aggregation and automated
  analysis of distributed systems traces.
\newblock In {\em Proceedings of the ACM Symposium on Cloud Computing}, pages
  76--91, 2021.

\bibitem{liu2020unsupervised}
Ping Liu, Haowen Xu, Qianyu Ouyang, Rui Jiao, Zhekang Chen, Shenglin Zhang,
  Jiahai Yang, Linlin Mo, Jice Zeng, Wenman Xue, et~al.
\newblock Unsupervised detection of microservice trace anomalies through
  service-level deep bayesian networks.
\newblock In {\em 2020 IEEE 31st International Symposium on Software
  Reliability Engineering (ISSRE)}, pages 48--58. IEEE, 2020.

\bibitem{nedelkoski2019anomaly}
Sasho Nedelkoski, Jorge Cardoso, and Odej Kao.
\newblock Anomaly detection from system tracing data using multimodal deep
  learning.
\newblock In {\em 2019 IEEE 12th International Conference on Cloud Computing
  (CLOUD)}, pages 179--186. IEEE, 2019.

\bibitem{bernieri1996line}
Andrea Bernieri, Giovanni Betta, and Consolatina Liguori.
\newblock On-line fault detection and diagnosis obtained by implementing neural
  algorithms on a digital signal processor.
\newblock {\em IEEE Transactions on Instrumentation and Measurement},
  45(5):894--899, 1996.

\bibitem{pearl2009causality}
Judea Pearl.
\newblock {\em Causality}.
\newblock Cambridge university press, 2009.

\bibitem{vaswani2017attention}
Ashish Vaswani, Noam Shazeer, Niki Parmar, Jakob Uszkoreit, Llion Jones,
  Aidan~N Gomez, {\L}ukasz Kaiser, and Illia Polosukhin.
\newblock Attention is all you need.
\newblock {\em Advances in neural information processing systems}, 30, 2017.

\bibitem{wang2022nenya}
Lu~Wang, Pu~Zhao, Chao Du, Chuan Luo, Mengna Su, Fangkai Yang, Yudong Liu,
  Qingwei Lin, Min Wang, Yingnong Dang, et~al.
\newblock Nenya: Cascade reinforcement learning for cost-aware failure
  mitigation at microsoft 365.
\newblock In {\em Proceedings of the 28th ACM SIGKDD Conference on Knowledge
  Discovery and Data Mining}, pages 4032--4040, 2022.

\bibitem{schulman2017proximal}
John Schulman, Filip Wolski, Prafulla Dhariwal, Alec Radford, and Oleg Klimov.
\newblock Proximal policy optimization algorithms.
\newblock {\em arXiv preprint arXiv:1707.06347}, 2017.

\bibitem{pearl1998graphical}
Judea Pearl.
\newblock Graphical models for probabilistic and causal reasoning.
\newblock {\em Quantified representation of uncertainty and imprecision}, pages
  367--389, 1998.

\bibitem{pearl2000models}
Judea Pearl et~al.
\newblock Models, reasoning and inference.
\newblock {\em Cambridge, UK: CambridgeUniversityPress}, 19(2), 2000.

\bibitem{shapley1953value}
Lloyd~S Shapley et~al.
\newblock A value for n-person games.
\newblock 1953.

\bibitem{paszke2019pytorch}
Adam Paszke, Sam Gross, Francisco Massa, Adam Lerer, James Bradbury, Gregory
  Chanan, Trevor Killeen, Zeming Lin, Natalia Gimelshein, Luca Antiga, et~al.
\newblock Pytorch: An imperative style, high-performance deep learning library.
\newblock {\em Advances in neural information processing systems}, 32, 2019.

\bibitem{dowhypaper}
Amit Sharma and Emre Kiciman.
\newblock Dowhy: An end-to-end library for causal inference.
\newblock {\em arXiv preprint arXiv:2011.04216}, 2020.

\bibitem{dowhy_gcm}
Patrick Bl{\"o}baum, Peter G{\"o}tz, Kailash Budhathoki, Atalanti~A.
  Mastakouri, and Dominik Janzing.
\newblock Dowhy-gcm: An extension of dowhy for causal inference in graphical
  causal models.
\newblock {\em arXiv preprint arXiv:2206.06821}, 2022.

\bibitem{wang2021groot}
Hanzhang Wang, Zhengkai Wu, Huai Jiang, Yichao Huang, Jiamu Wang, Selcuk Kopru,
  and Tao Xie.
\newblock Groot: An event-graph-based approach for root cause analysis in
  industrial settings.
\newblock In {\em 2021 36th IEEE/ACM International Conference on Automated
  Software Engineering (ASE)}, pages 419--429. IEEE, 2021.

\bibitem{sanderson2010christopher}
Mark Sanderson.
\newblock Christopher d. manning, prabhakar raghavan, hinrich sch{\"u}tze,
  introduction to information retrieval, cambridge university press. 2008.
  isbn-13 978-0-521-86571-5, xxi+ 482 pages.
\newblock {\em Natural Language Engineering}, 16(1):100--103, 2010.

\bibitem{zoph2016neural}
Barret Zoph and Quoc~V Le.
\newblock Neural architecture search with reinforcement learning.
\newblock {\em arXiv preprint arXiv:1611.01578}, 2016.

\bibitem{he2021automl}
Xin He, Kaiyong Zhao, and Xiaowen Chu.
\newblock Automl: A survey of the state-of-the-art.
\newblock {\em Knowledge-Based Systems}, 212:106622, 2021.

\bibitem{pham2018efficient}
Hieu Pham, Melody Guan, Barret Zoph, Quoc Le, and Jeff Dean.
\newblock Efficient neural architecture search via parameters sharing.
\newblock In {\em International conference on machine learning}, pages
  4095--4104. PMLR, 2018.

\bibitem{pu2021learning}
Xingyue Pu, Tianyue Cao, Xiaoyun Zhang, Xiaowen Dong, and Siheng Chen.
\newblock Learning to learn graph topologies.
\newblock {\em Advances in Neural Information Processing Systems},
  34:4249--4262, 2021.

\bibitem{munikoti2022challenges}
Sai Munikoti, Deepesh Agarwal, Laya Das, Mahantesh Halappanavar, and
  Balasubramaniam Natarajan.
\newblock Challenges and opportunities in deep reinforcement learning with
  graph neural networks: A comprehensive review of algorithms and applications.
\newblock {\em arXiv preprint arXiv:2206.07922}, 2022.

\bibitem{li2023network}
Zhuoran Li, Xing Wang, Ling Pan, Lin Zhu, Zhendong Wang, Junlan Feng, Chao
  Deng, and Longbo Huang.
\newblock Network topology optimization via deep reinforcement learning.
\newblock {\em IEEE Transactions on Communications}, 2023.

\bibitem{meirom2021controlling}
Eli Meirom, Haggai Maron, Shie Mannor, and Gal Chechik.
\newblock Controlling graph dynamics with reinforcement learning and graph
  neural networks.
\newblock In {\em International Conference on Machine Learning}, pages
  7565--7577. PMLR, 2021.

\bibitem{alipour2022multiagent}
Mir~Mohammad Alipour and Mohsen Abdolhosseinzadeh.
\newblock A multiagent reinforcement learning algorithm to solve the community
  detection problem.
\newblock {\em Signal and Data Processing}, 19(1):87--100, 2022.

\bibitem{chen2021contingency}
Haipeng Chen, Wei Qiu, Han-Ching Ou, Bo~An, and Milind Tambe.
\newblock Contingency-aware influence maximization: A reinforcement learning
  approach.
\newblock In {\em Uncertainty in Artificial Intelligence}, pages 1535--1545.
  PMLR, 2021.

\bibitem{li2021practical}
Zeyan Li, Junjie Chen, Rui Jiao, Nengwen Zhao, Zhijun Wang, Shuwei Zhang,
  Yanjun Wu, Long Jiang, Leiqin Yan, Zikai Wang, et~al.
\newblock Practical root cause localization for microservice systems via trace
  analysis.
\newblock In {\em 2021 IEEE/ACM 29th International Symposium on Quality of
  Service (IWQOS)}, pages 1--10. IEEE, 2021.

\bibitem{li2022mining}
Mingjie Li, Minghua Ma, Xiaohui Nie, Kanglin Yin, Li~Cao, Xidao Wen, Zhiyun
  Yuan, Duogang Wu, Guoying Li, Wei Liu, et~al.
\newblock Mining fluctuation propagation graph among time series with active
  learning.
\newblock In {\em Database and Expert Systems Applications: 33rd International
  Conference, DEXA 2022, Vienna, Austria, August 22--24, 2022, Proceedings,
  Part I}, pages 220--233. Springer, 2022.

\bibitem{kalisch2007estimating}
Markus Kalisch and Peter B{\"u}hlman.
\newblock Estimating high-dimensional directed acyclic graphs with the
  pc-algorithm.
\newblock {\em Journal of Machine Learning Research}, 8(3), 2007.

\bibitem{ma2020automap}
Meng Ma, Jingmin Xu, Yuan Wang, Pengfei Chen, Zonghua Zhang, and Ping Wang.
\newblock Automap: Diagnose your microservice-based web applications
  automatically.
\newblock In {\em Proceedings of The Web Conference 2020}, pages 246--258,
  2020.

\bibitem{chen2014causeinfer}
Pengfei Chen, Yong Qi, Pengfei Zheng, and Di~Hou.
\newblock Causeinfer: Automatic and distributed performance diagnosis with
  hierarchical causality graph in large distributed systems.
\newblock In {\em IEEE INFOCOM 2014-IEEE Conference on Computer
  Communications}, pages 1887--1895. IEEE, 2014.

\bibitem{guo2020survey}
Ruocheng Guo, Lu~Cheng, Jundong Li, P~Richard Hahn, and Huan Liu.
\newblock A survey of learning causality with data: Problems and methods.
\newblock {\em ACM Computing Surveys (CSUR)}, 53(4):1--37, 2020.

\end{thebibliography}









\end{document}